\documentclass[runningheads]{llncs}
\usepackage[T1]{fontenc}
\usepackage{graphicx}
\usepackage{amsmath,graphicx}
\usepackage{multirow}
\usepackage{booktabs}
\usepackage{makecell}
\usepackage{xcolor}

\begin{document}
\title{FixPix: \underline{Fix}ing Bad \underline{Pix}els using Deep Learning\thanks{Supported by Samsung}}
%
%
\author{Sreetama Sarkar \and
Xinan Ye \and
Gourav Datta \and
Peter Beerel}

\authorrunning{S. Sarkar et al.}
%
\institute{University of Southern California, Los Angeles California 90089, USA \\
\email{\{sreetama,xinanye,gdatta,pabeerel\}@usc.edu}}
%
\maketitle              
\begin{abstract}
Efficient and effective on-line detection and correction of bad-pixels can improve yield and increase the expected lifetime of image sensors. This paper presents a comprehensive Deep Learning (DL) based on-line detection and correction approach, suitable for a wide range of pixel corruption rates. 
A confidence calibrated segmentation approach is introduced, which achieves nearly perfect bad pixel detection, even with a few training samples. A computationally light-weight correction algorithm is proposed for low rates of pixel corruption, that surpasses the accuracy of traditional interpolation-based techniques. In addition, a vision transformer (ViT) auto-encoder based image reconstruction approach is presented which 
yields promising results for high rates of pixel corruption or clustered defects. Unlike previous methods, which use proprietary images, we demonstrate the efficacy of the proposed methods on the open-source Samsung S7 ISP and MIT-Adobe FiveK datasets. Our approaches yield up to 99.6\% detection accuracy with ${<}0.6\%$ false positives and corrected images within 1.5\% average pixel error from 70\% corrupted images. 
We achieve correction error at par with the state-of-the-art (SoTA) DL methods for clustered defects with less than half the computational cost.

\end{abstract}
\section{Introduction}
\label{sec:intro}

There have been remarkable technological advances in the development of CMOS image sensors with improvement in quality, efficiency, and fault-tolerance \cite{scaling_cis}. Nevertheless, pixel defects can occur in these sensors during the manufacturing process or later during operation, are permanent, and increase in number over the lifetime of the sensor. These defects degrade the sensor yield and effectiveness and consequently increase cost.  Pixel defects are important for a wide range of image sensors, but particularly for sensors that are regularly exposed to high levels of light, electrical energy, or radiation, such as in satellites and telescopes, which leads to high rates of pixel corruption.  

Traditionally, pixel defects are detected only during manufacturing \cite{lynx}. However, the resulting static pixel defect maps do not capture defects developed during the lifetime of the sensor. Online pixel defect detection usually relies either on the analysis of neighborhood pixels within the current frame \cite{pinto2012dynamic,kakarala2004bad} or on multiple frames \cite{leung2008automatic,mijatovic2012implementation}, \textcolor{black}{rendering them useless when pixel defects occur in nearby pixels or clusters.}
In this work, we propose bad pixel detection leveraging multiple frames capable of detecting clustered defects without the need to store pixel information from individual frames, thereby eliminating any increased memory overhead while achieving perfect detection.

The detected pixel defects are typically corrected using interpolation algorithms, such as nearest neighbor interpolation \cite{nn}, linear filtering \cite{linear}, and median filtering \cite{Wang2009AdaptiveDC}. \textcolor{black}{Traditional approaches are heuristic-based and often tailored for a particular sensor type and error pattern. Clustered defect correction algorithms \cite{Schberl2011SparsitybasedDP,Lee2021UsingDL} assume that the defect locations are already known, whereas, defect detection and correction in commercial image signal processors (ISP) are not equipped to detect or correct clustered defects. Sophisticated approaches like adaptive filtering \cite{Schberl2011SparsitybasedDP} aim to estimate edges and directions and are extremely complicated and harder to optimize. This motivates a learning-based method that is applicable to a wide range of error patterns, error rates and sensor types.} Motivated by successes in a wide variety applications, deep learning (DL) have also been explored in the area of pixel defect detection and correction \cite{Kalyanasundaram2020APA,Lee2021UsingDL}.

In this paper, we propose DL based online bad pixel detection and correction \textcolor{black}{on Bayer images} suitable for both photographic and computer vision (CV) applications. Our goal is to improve sensor yields during manufacturing as well as increase their effective lifetime. More specifically, we first propose to detect bad pixels, which gives us the error rate in the image. We then propose two different strategies for correcting low and high rates of pixel corruption. For low error rates, we propose a lightweight patch-based pixel correction on extracted patches around the detected bad pixel. For very high error rates and clustered defects, we propose a ML-based complete reconstruction algorithm. \textcolor{black}{We demonstrate results by injecting errors on two different datasets, Samsung S7 ISP \cite{Schwartz2018DeepISPTL} and MIT-Adobe FiveK dataset \cite{fivek} that have RAW Bayer CFA format images. Our approach for detecting and correcting image errors can be easily extended to all types of images, including grayscale, RGB, or IR images.} 


\noindent\textbf{Contributions} Our contributions can be summarized as follows. (1) We propose a binary segmentation method for effective detection of bad pixels. While this approach achieves nearly perfect detection for large datasets, the detection rate drops for smaller datasets. To mitigate this gap, we propose confidence calibration using multiple images during inference. Our confidence-calibrated segmentation approach yields an improvement of up to 20\% over regular binary segmentation. (2) We propose a lightweight patch based pixel correction using multi-layer perceptron (MLP) models for low error rates, that outperforms existing interpolation techniques. More specifically, our MLP model exceeds reported values for Adaptive Defect Correction \cite{Tanbakuchi2003AdaptivePD} by 7.05dB and linear \cite{linear} and median \cite{Wang2009AdaptiveDC} interpolation by 4.85dB, for the same error rate. (3) For extremely high rates of pixel corruption, we propose a fail-safe autoencoder based image reconstruction approach, that needs no prior detection. This approach achieves a Normalized Mean Squared Error (NMSE) of 1.55\% for up to 70\% corrupted pixels. For $5\times5$ defect clusters, it achieves an NMSE as low as 0.3\%, which is at par with SoTA DL approaches \cite{Lee2021UsingDL}, with less than half the parameters and computational cost.

\section{Background}
\label{sec:background}
\textbf{Bad Pixel Detection:} 
There are two main types of bad pixel detection methods: online and offline. Offline methods detect bad pixels during the manufacturing process, while online methods are used to detect defects throughout the sensor's lifetime. Traditionally, offline detection involves observing which pixel values remain unchanged across images and creating a map of defective pixels \cite{lynx}. This map is then stored in a non-volatile memory integrated with the sensor chip to guide downstream pixel correction logic \cite{lynx}. Online detection is crucial for identifying defects during the lifetime of a sensor, and can be done by analyzing either a single frame or multiple frames. Single-frame detection \cite{pinto2012dynamic,kakarala2004bad,chan2009dead,cho2011real} involves comparing the values of the pixel in question to those of its neighboring pixels. Two commonly used defect detection and correction algorithms in commercial ISP are Pinto \cite{pinto2012dynamic} and Kakarala \cite{kakarala2004bad}. Pinto \cite{pinto2012dynamic} uses a $3\times3$ neighborhood and identifies a pixel as defective if it has the highest or lowest value in the neighborhood. Naturally, this method cannot detect multiple defective pixels in the same neighborhood. \cite{kakarala2004bad,chan2009dead,cho2011real} set upper and lower thresholds based on values of neighboring pixels, and any pixel that falls outside of this range is flagged as defective. 
\cite{ElYamany2017RobustDP} performs rule-based analysis of pixel deviation from local average estimates of same color neighbors. None of these methods are equipped to detect clustered defects or multiple bad pixels in close vicinity. \cite{mijatovic2012implementation,leung2008automatic,tajbakhsh2011efficient} uses multi-frame processing to detect defects. \cite{mijatovic2012implementation} uses a combination of neighborhood pixels and temporal consistency. 
The pixels exceeding the average neighborhood pixel values by a pre-defined threshold are stored as candidate bad pixels and monitored over a number of time steps. Pixels whose value remains unchanged over a sufficient time period are declared defective. \cite{leung2008automatic} uses Bayesian statistics of image sequences collected over days for defect detection. These methods incur a large memory overhead for storing image statistics.
More recently, deep learning has also been used for bad pixel detection \cite{ye2018intelligent,Kalyanasundaram2020APA,ma201941}. Kalyanasundaram et al. \cite{Kalyanasundaram2020APA} proposed a MLP model for detection of isolated defects, although it needs some initial pre-processing steps. \cite{ye2018intelligent} uses convolutional neural networks (CNNs) while \cite{ma201941} uses a YOLOv3 \cite{redmon2016you} based architecture for defect detection.

\vspace{2mm}
\noindent\textbf{Bad Pixel Correction:}
Pixel defect correction is typically performed using interpolation. While nearest neighbor interpolation \cite{nn} replaces the defective pixel with its nearest non-defective pixel value in 2D space, linear filtering \cite{linear}, and median filtering \cite{Wang2009AdaptiveDC} compute the mean and median of a few non-defective neighboring pixels to replace the defective pixel. Bad pixel correction in Kakarala \cite{kakarala2004bad} and Pinto \cite{pinto2012dynamic} are performed using linear and median filtering, respectively, which leads to image blurring when edges are present in the patch. More advanced interpolation techniques such as Adaptive Defect Correction (ADC) \cite{Tanbakuchi2003AdaptivePD} tries to estimate edges and directions for defect correction although it can only work on Bayer pattern images. Sparsity-based Defect Interpolation \cite{Schberl2011SparsitybasedDP} devises a sparsity based high-complexity iterative algorithm that  leverages complex-valued frequency selective extrapolation and outperforms previous interpolation techniques. These correction approaches assume that defect locations are already known during manufacturing. DL approaches have also been explored for pixel defect corrections \cite{chen2023hisp,Lee2021UsingDL}. For pixel-defect correction of flat-panel radiography images, \cite{Lee2021UsingDL} uses different DL approaches: a single-layer ANN, a multi-layer CNN, a concatenated CNN, and GANs, and infer that their concatenated CNN performs the best for correcting clustered defects.

\section{Bad Pixel Detection}
\label{sec:detection}

\begin{figure}[htbp]
    \centering
      \includegraphics[width=0.6\textwidth]{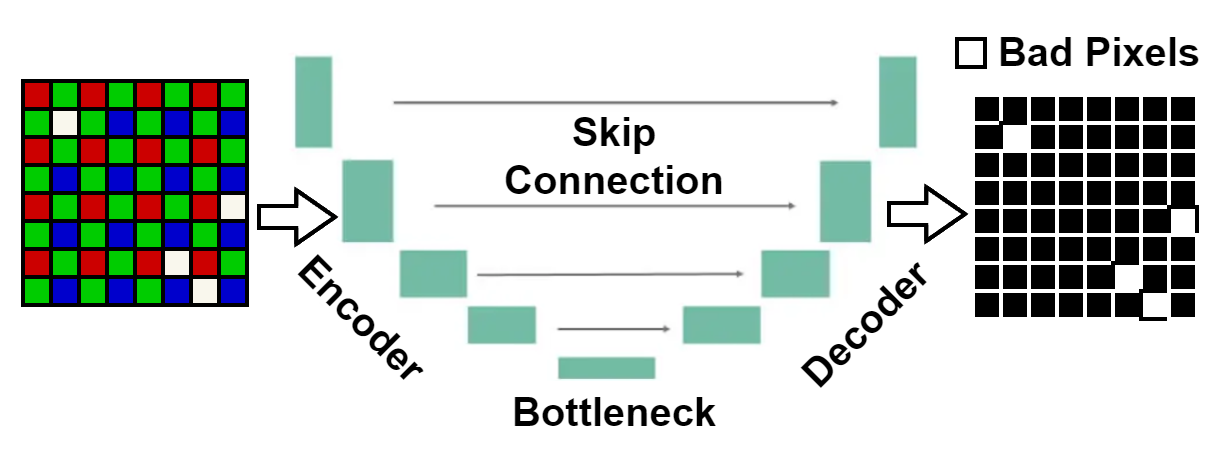}
    \caption{Bad pixel detection using binary segmentation}
    \label{fig:bp_detect}
    \vspace{-4mm}
\end{figure}

Semantic segmentation \cite{long2015fully,Ronneberger2015UNetCN,chen2017deeplab,biomed_segmentation} is a common task in computer vision, where each pixel in an image is assigned a specific class. Segmentation is popular for understanding image context in applications like autonomous driving \cite{Geiger2012CVPR} or medical image analysis \cite{Ronneberger2015UNetCN}. We formulate bad pixel detection as a binary segmentation problem consisting of two classes: good pixels and bad pixels (Figure \ref{fig:bp_detect}). We perform detection using U-Net \cite{Ronneberger2015UNetCN}, originally proposed for medical image segmentation. It has an encoder-decoder architecture making the network U-shaped (as shown in figure \ref{fig:bp_detect}), where the encoder consists of a set of downsampling layers while the decoder consists of a set of upsampling layers. \textcolor{black}{The model takes single channel Bayer images with defective pixels as input and generates a binary map indicating good and bad pixels as output, which has the same dimension as the input. This method can be extended for RGB inputs by training a U-Net model with three input channels instead of one input channel.} 
The model is trained using a combination of binary cross-entropy and dice loss.

\begin{figure}[htbp]
    \centering
      \includegraphics[width=0.6\textwidth]{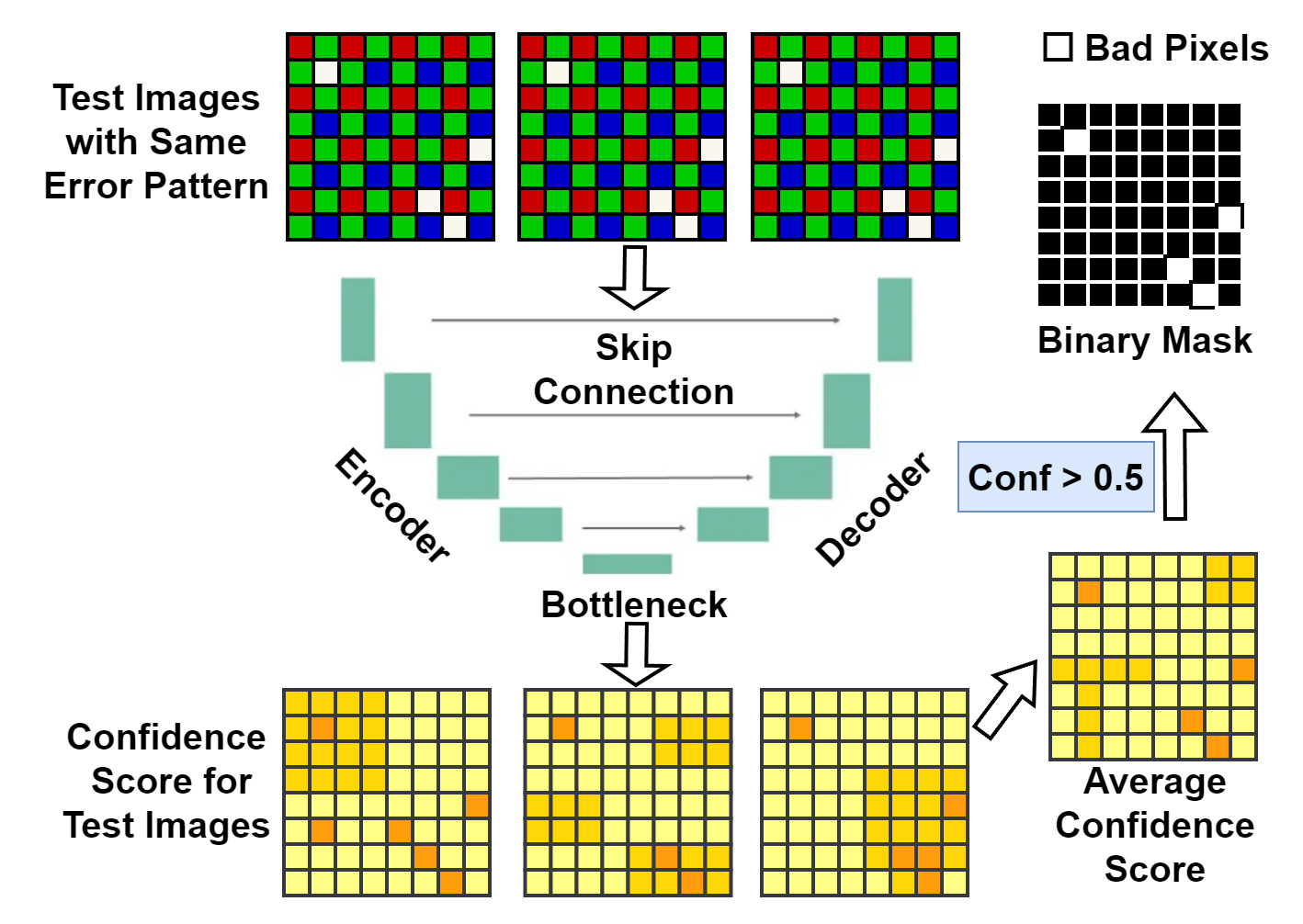}
    \caption{Bad pixel detection using confidence-calibrated segmentation}
    \label{fig:bpdcc}
\end{figure}

However, simply using binary segmentation cannot achieve perfect detection, particularly when there are limited number of images for training the segmentation model (see Section \ref{sec:results_detection}).
Due to nature of defects, the corrupted pixels always occur in the same location across images. A single image may not be enough to correctly identify all bad pixel locations. We leverage the predictions from multiple test images for more reliable bad pixel detection. For semantic segmentation, the model outputs a set of probability or confidence scores indicating if a pixel belongs to a particular class. Instead of taking probability values from a single image, we take mean probability score of $n$ images during test time, which is then thresholded to obtain final class labels, as shown in Figure \ref{fig:bpdcc}. This approach is termed as confidence-calibrated segmentation. Our results show significant improvement in detection performance (see Section \ref{sec:results_detection}). 
\textcolor{black}{Notably, we maintain a pixel-wise cumulated probability score (sum of probability scores) across multiple frames, instead of maintaining individual pixel-wise probabilities for each frame. Therefore, we just need to store the information corresponding to the number of pixels in a single frame, making our method more memory-efficient as compared to existing multi-frame detection approaches.}

\section{Bad Pixel Correction}
\label{sec:correction}
For correction of bad pixels, we propose two different approaches to deal with different error rates. The error rate can be measured using the detection method proposed above. First, we propose a patch-based correction approach, where a $n\times n$ patch around the detected bad pixel is extracted, and passed through the correction network to obtain the actual value of the erroneous central pixel. While this method performs reasonably well for low error rates, it fails when the bad pixels are clustered or the number of bad pixels in a patch is very high. For this, we propose a fail-safe, a Vision Transformer based Autoencoder (ViT AE) \cite{dosovitskiy2010image,he2022masked} for pixel correction using complete image reconstruction.

\begin{figure}[htb]

%
\begin{minipage}[b]{.48\linewidth}
  \centering
  \centerline{\includegraphics[width=0.8\linewidth]{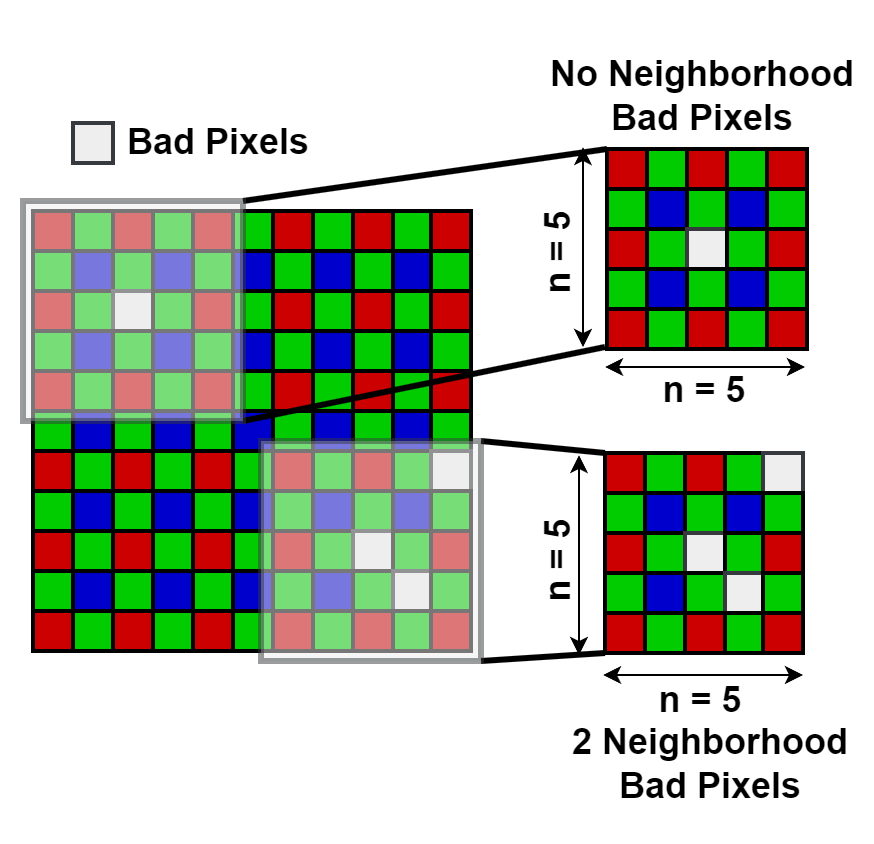}}
  \centerline{(a) Extracting 5$\times$5 patches}\medskip
    \label{fig:extract_patch}
\end{minipage}
\hfill
\begin{minipage}[b]{0.48\linewidth}
  \centering
  \centerline{\includegraphics[width=0.8\linewidth]{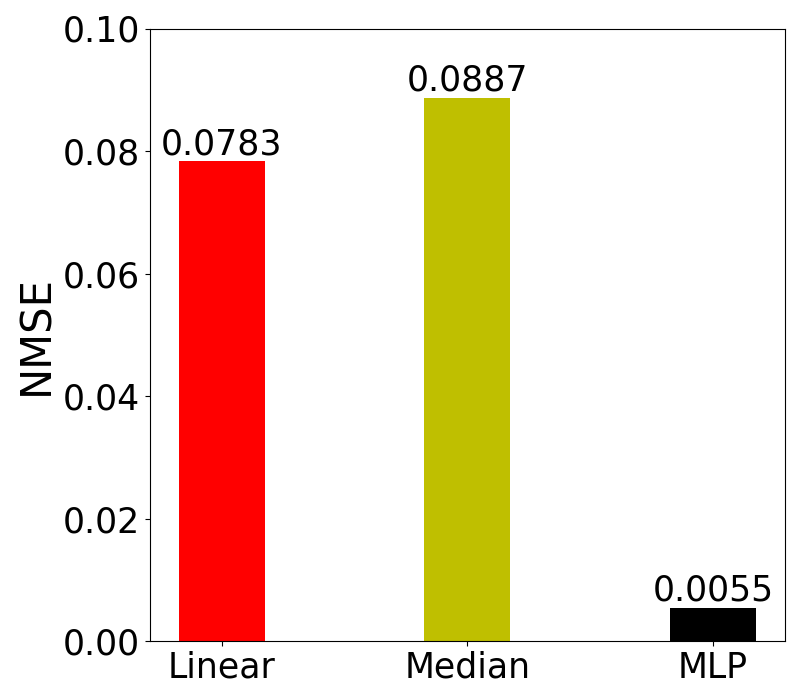}}
  \centerline{(b) NMSE comparison}\medskip
  \label{fig:mean_median_comparison}
\end{minipage}
\vspace{-0.5cm}
\caption{Patch based bad pixel correction}
\vspace{-0.6cm}
\label{fig:res}
\end{figure}

\vspace{5mm}
\noindent\textbf{MLP based Correction:}
While bad pixel detection needs to be applied periodically during the lifetime of the sensor (e.g., during the boot-up process), bad pixel correction has to be performed on every single captured image. Hence, the correction algorithm should be preferably lightweight. We build a 2-layer MLP, consisting of 2 fully-connected layers with ReLU activation, to predict the central pixel from neighboring pixel values. The ReLU layers introduce non-linearity, which helps the network better estimate the optimal pixel value and outperform traditional interpolation approaches. 
We compare our approach with linear \cite{linear} and  median \cite{Wang2009AdaptiveDC} filtering (Figure \ref{fig:mean_median_comparison}) and observe that our approach yields $\sim14.2\times$ lower NMSE than these methods. 


A $n\times n$ patch may contain multiple bad pixels in the neighborhood of the central pixel, as shown in Figure \ref{fig:extract_patch}, which makes the problem of pixel correction harder. We adopt two different approaches to mitigate this problem: \textit{increasing patch size} and \textit{training models with neighborhood bad pixels}. Increasing patch size provides the model with a larger window of neighborhood pixels for prediction of the central pixel, which is particularly advantageous when there are multiple bad pixels in close vicinity. On the other hand, training with corrupted pixels imparts the ability to infer correct prediction discarding defective pixels in the neighborhood. While very effective for low error rates, for high levels of image corruption, these methods experience notable deterioration (see Section \ref{sec:results_detection}), motivating a secondary approach.

\vspace{5mm}
\noindent\textbf{Image Reconstruction using a ViT AE:}
An Autoencoder (AE) \cite{denoisingAE,an2015variational} has an encoder-decoder architecture, where the encoder learns the latent features from input images and the decoder reconstructs the image using those latent features. They are used for a wide range of vision tasks, including anomaly detection \cite{zhou2017anomaly,zhao2017anomaly,an2015variational}, segmentation \cite{Ronneberger2015UNetCN,chen2017deeplab} and super-resolution~\cite{dong2015imagesuperr,mao2016imagesuperr}. Denoising autoencoders (DAE) \cite{denoisingAE} or masked autoencoders (MAE) \cite{he2022masked} are used as pre-training for very large models. While DAE injects noise, MAE masks out large portions of the input, and the model learns to reconstruct the image from partial input information, learning robust features. 

Unlike these approaches, we use AE with the goal to recover original pixel values from corrupted images. We design a ViT based AE inspired from MAE \cite{he2022masked}, \textcolor{black}{which takes corrupted single-channel Bayer images as input}. However, differing from \cite{he2022masked}, we do not mask image portions or use mask tokens, but input all embeddings from the corrupted images into the encoder blocks. The AE is trained by minimizing normalized error on the corrupted pixels. We demonstrate that this method, although computationally expensive compared to an MLP and hence unnecessary for low error rates, yields significant benefits for high rates of pixel corruption. More importantly, this method does not require exact bad pixel locations. Therefore, there is no need to perform detection every single time prior to correction, thereby saving detection cost. Moreover, the size of the AE model scales with input size, meaning, based on the model size that can fit into the sensor chip, we can break the input image into patches and perform patch wise reconstruction. For an input of $15\times15$, we demonstrate results using an AE model with only 2 encoder and decoder layers consisting of only 11K parameters (Table \ref{tab:concatcnn_comparison}). 

\section{Experimental Results}
\label{sec:exp}
\subsection{Experimental Setup}
\textbf{Models and Dataset:} Our approaches are evaluated on the Samsung S7 ISP \cite{Schwartz2018DeepISPTL} and the Canon EOS 5D subset of the MIT-Adobe FiveK dataset \cite{fivek} datasets. The S7 ISP is a small dataset, consisting of 110 image pairs, captured using the Samsung S7 rear camera, whereas the MIT FiveK is a large-scale dataset, containing 5,000 photographs taken with SLR cameras, from which we extract 777 Canon images, similar to \cite{xing21invertible}. We consider the raw images for this task and inject bad pixels to the images to evaluate our approaches. The datasets are split into train, validation, and test sets in a ratio of 8:1:1. Detection is performed using U-Net \cite{Ronneberger2015UNetCN} segmentation model, and correction is performed using a 2-layer MLP \cite{lecun2015deep} and ViT AE, inspired from \cite{he2022masked}.

\vspace{2mm}
\noindent\textbf{Bad pixel injection:}    
Pixel defects in image sensors have different types. While \textit{dead pixels} are permanently stuck at 0, \textit{hot pixels} or \textit{stuck pixels} maybe permanently bright. Pixel defects may also cause them to deviate from their original value. In our framework, the bad pixel value is obtained by adding at least $\pm \delta$ variation to original pixel value, but still within the permissible range of pixel values. We test our approach over a wide range of $\delta$. 
The lower the deviation from its original value, the harder it is to detect a bad pixel, whereas, higher the deviation in neighboring pixels, harder it is for correction. 
The number of bad pixels expected is determined by manufacturing facilities as well as 
sensor lifetime models. We test over a wide range of error rates from 0.01\% to 85\% with 
the location of bad pixels selected at random. 

\vspace{2mm}
\noindent\textbf{Training Framework:}
Both U-Net and ViT AE models are trained for 50 epochs on S7 ISP and 10 epochs on MIT FiveK datasets. For U-Net, we use a step learning rate (lr), starting with a lr of 0.001 and decaying by 0.5 every 10 epochs. ViT AE is trained with an initial lr of 0.01, a linear increase in lr for the first 5 epochs, and cosine decay in lr for the rest of the training epochs. The MLP models for correction are trained for 50 epochs using a learning rate of 0.01. Since raw images have very high dimension ($\sim$3000$\times$4000), each image is broken down into 64 patches before being fed into U-Net or ViT AE, to reduce computation. They can be decomposed into even smaller patches based on computational constraints.

\vspace{2mm}
\noindent\textbf{Evaluation metrics:}
The detection approach is evaluated using \textit{Precision} and \textit{Recall}. \textit{Precision} is defined as $\text{TP}/(\text{TP}+\text{FP})$ and \textit{Recall} is defined as $\text{TP}/(\text{TP}+\text{FN})$ where $\text{TP}$, $\text{FP}$, and $\text{FN}$ refer to true positives, false positives, and false negatives, respectively. In this case, bad pixels are considered positives. Thus, \textit{recall} quantifies the detection rate and \textit{precision} quantifies the false positive ratio. The correction approach is evaluated using NMSE given by $\frac{\lVert p_{pred}-p_{act}\rVert_2^2}{\lVert p_{act}\rVert_2^2}$ where $p_{act}$ and $p_{pred}$ refer to actual and predicted pixel values. PSNR or peak signal to noise ratio is used to measure the quality of the corrected image with respect to the original image. PSNR is given by $10log_{10}(\frac{1}{MSE})$ where MSE is the mean squared error between original and corrected image.

\subsection{Results and Analysis}
\label{sec:results_detection}

\begin{figure}[htb]
\begin{minipage}[b]{.48\linewidth}
  \centering
  \centerline{\includegraphics[width=0.9\linewidth]{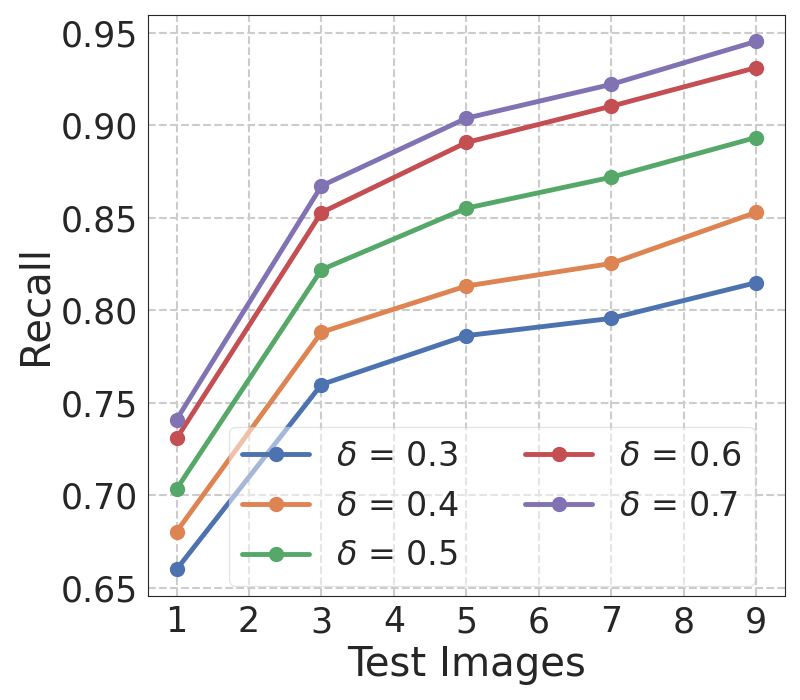}}
  \centerline{(a) Recall}\medskip
\end{minipage}
\hfill
\begin{minipage}[b]{0.48\linewidth}
  \centering
  \centerline{\includegraphics[width=0.9\linewidth]{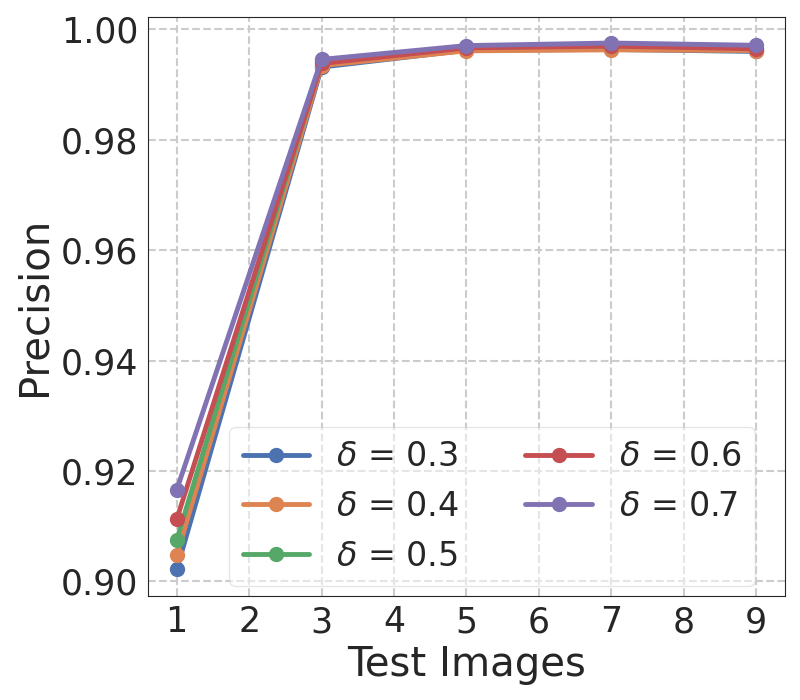}}
  \centerline{(b) Precision}\medskip
\end{minipage}
\vspace{-0.5cm}
\caption{Precision and recall vs \# of test images for confidence calibration on S7 ISP dataset (error rate=0.01\%)}
\label{fig:det_graph}
\end{figure}

Table \ref{tab:results1} summarizes the results for bad pixel detection and correction using the proposed approaches for error rates ranging from 0.01\% to 70\% and bad pixel values deviating from the original pixel value by 70\%. Detection results are reported for a single test image. While for the larger dataset MIT FiveK, we are able to achieve 99.6\% detection accuracy, even with a single test image, the obtained detection rate is lower for the smaller S7 ISP dataset. 
\textcolor{black}{More specifically, we observe a lower recall or detection rate for an error rate of 0.01\%. The number of bad pixels in the training set for an error rate of 0.01\%, is much smaller than the number of good pixels, resulting in a skewed distribution for the binary segmentation task, which probably leads to poor training and consequently, a performance drop.} To mitigate this, we leverage prediction confidence for multiple test images (see Figure \ref{fig:det_graph}). NMSE values are reported for both patch-based correction using an MLP (NMSE$_{MLP}$) and image reconstruction using an ViT (NMSE$_{AE}$). The MLP model is applied on 5$\times$5 patches surrounding the pixel to be corrected, whereas, the AE is applied on the entire image, both having the specified error rate. While patch-based pixel correction is effective for low error rates, it suffers up to 31\% pixel error when 70\% pixels are bad. The AE model successfully reduces this error to 1.55\%.

\begin{table}[h]
\centering
\begin{tabular}{c|c|c|c|c|c}
\toprule
\textbf{Dataset} & \textbf{Error} & \multicolumn{2}{c|}{\textbf{Detection}} & \multicolumn{2}{c}{\textbf{Correction}} \\
 & (\%) & \textbf{Recall} & \textbf{Precision} & \textbf{NMSE$_{MLP}$} & \textbf{NMSE$_{AE}$}\\
\midrule
\multirow{2}{*}{\rotatebox{0}{\makecell{S7 ISP}}} & 0.01 & 0.85 & 0.96 & 0.005 & 0.053\\
& 70 & 0.95 & 0.99 & 0.26 & 0.098\\
\midrule
\multirow{2}{*}{\rotatebox{0}{\makecell{MIT FiveK}}} & 0.01 & 0.996 & 0.994 & 0.0009 & 0.0036\\
& 70 & 0.994 & 0.995 & 0.31 & 0.0155\\
\bottomrule
\end{tabular}
\vspace{0.2cm}
\caption{Detection and correction results for widely different error rates (with $\delta$=0.7).}
\label{tab:results1}
\end{table}
\noindent\textbf{Improving Detection using Confidence Calibration:}
In Table \ref{tab:results1}, we observe that for the S7 ISP dataset, we are not able to detect nearly 15\% of the bad pixels, even for a $\delta$ variation of 0.7. To address this, we use our confidence calibration approach. Figure \ref{fig:det_graph} illustrates precision and recall values for a fixed error pattern and different $\delta$ variations, with an increase in the number of images used during inference. A lower $\delta$ variation from the original pixel value makes it harder to detect the bad pixel, resulting in lower recall. The increasing trend in precision and recall with increased number of images used during inference reaffirms our hypothesis that using multiple images during test time helps in more reliable pixel detection. We observe that using 9 test images, the maximum number supported by our test set, yields an improvement of $\sim$20\% in detection rate, compared with a single test image. Note, the recall values in Table \ref{tab:results1} and Figure \ref{fig:det_graph} are different for same $\delta$ and error rate. This is because the difference in the injected error pattern or the location of the injected bad pixels. 




\begin{figure}[htbp]
\begin{minipage}[b]{.48\linewidth}
  \centering
  \centerline{\includegraphics[width=0.9\linewidth]{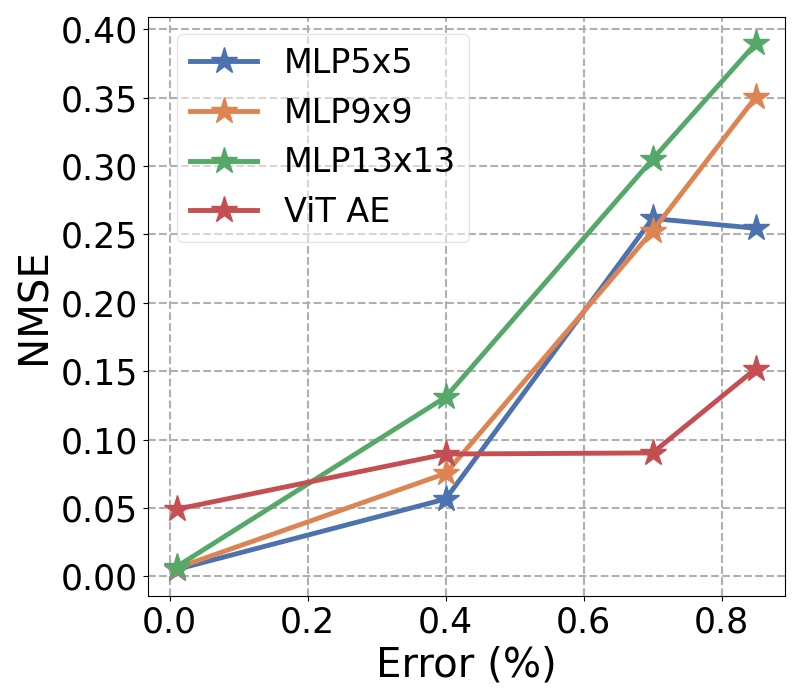}}
  \centerline{(a) Samsung S7 ISP}\medskip
\end{minipage}
\hfill
\begin{minipage}[b]{0.48\linewidth}
  \centering
  \centerline{\includegraphics[width=0.9\linewidth]{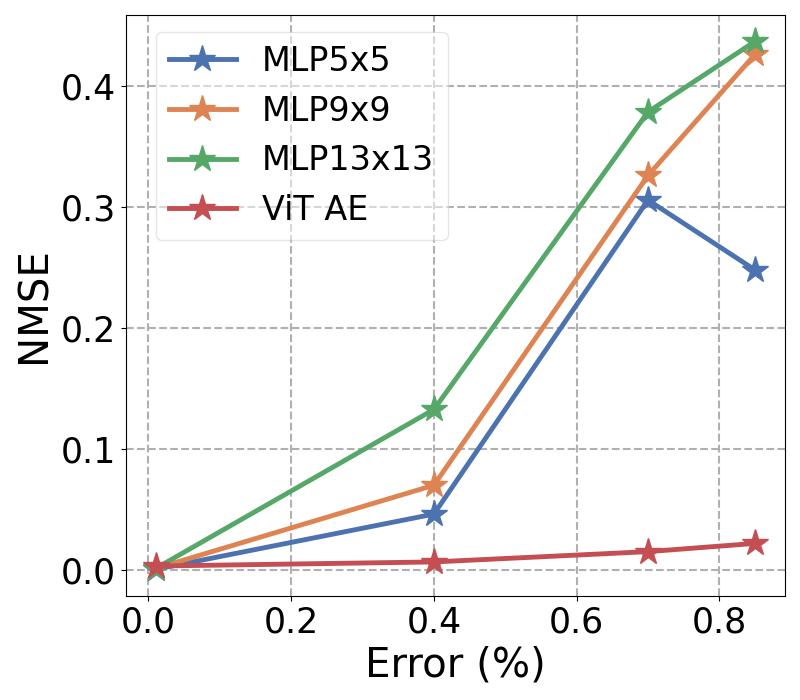}}
  \centerline{(b) MIT Adobe 5K}\medskip
  
\end{minipage}
\vspace{-0.5cm}
\caption{Comparison of ViT AE and MLP based correction with S7 ISP and MIT FiveK datasets with a wide range of error rates}
\label{fig:mae_vs_mlp}
\vspace{-0.5cm}
\end{figure}

\vspace{2mm}
\noindent\textbf{Correction using MLP vs ViT AE:}
In Figure \ref{fig:mae_vs_mlp}, we compare patch based pixel correction vs AE based reconstruction for different error rates. MLP models are trained with varying patch sizes with pre-defined error rates. We observe that increasing patch size for MLP models is ineffective when the number of bad pixels scales with patch size. 
ViT, however, learns from the global image context and maintains a low NMSE even for very high error rates, achieving 2.2\% NMSE for 85\% corrupted pixels. While ViT AE performs significantly better for high error rates, patch wise detection and correction is more effective for low error rates. \textcolor{black}{This is primarily because local context is more helpful if the neighborhood pixels are not corrupted. However, if there are too many bad pixels in the neighborhood as in case of clustered defects, the global context is more effective in pixel correction.} From Figure \ref{fig:mae_vs_mlp}, we observe that the MLP still performs better for an error rate of 40\% on the S7-ISP dataset, while it suffers a small increase on MIT Adobe 5K. Therefore, we set an error rate of 40\% as the threshold for switching to AE based reconstruction.  


\begin{figure}[htbp]
    \centering
    \includegraphics[width=0.55\textwidth]{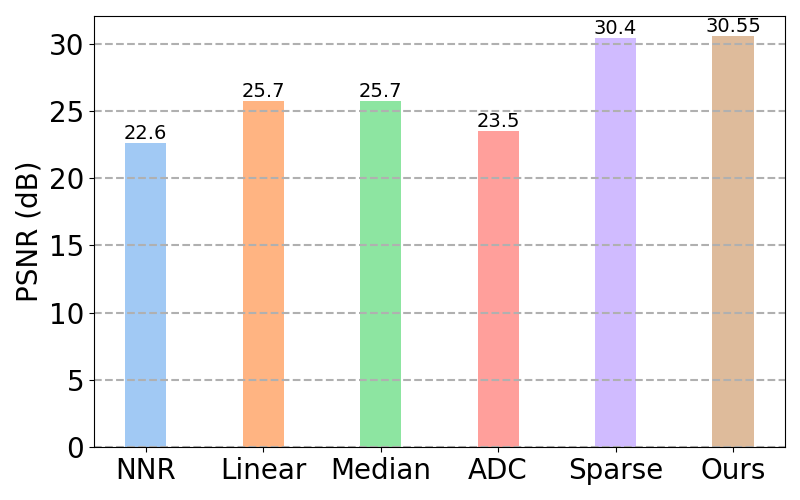}
    \caption{Comparison of MLP based correction with existing interpolation methods}
    \label{tab:interpolation}
\end{figure}

\noindent\textbf{Comparison with SoTA Interpolation Methods:}
We compare MLP based pixel correction with existing interpolation techniques for an error rate of 20\%, where 5 erroneous pixels are present in a 5$\times$5 patch. Our MLP model achieves a PSNR of 30.55 dB on the S7 ISP dataset, which is higher than reported for all interpolation techniques described in Section \ref{sec:background}. More specifically, a comparison with the results reported in Table \ref{tab:interpolation} suggests that our model exceeds ADC \cite{Tanbakuchi2003AdaptivePD} by 7.05dB and linear \cite{linear} and median \cite{Wang2009AdaptiveDC} interpolation by 4.85dB and Sparsity-based Defect Interpolation \cite{Schberl2011SparsitybasedDP} by 0.15dB. The numbers for the interpolation methods are taken from \cite{Schberl2011SparsitybasedDP}. Thus, simple learning-based techniques is found to outperform complex rule-based handcrafted interpolation methods.

\vspace{2mm}
\noindent\textbf{Comparison with SoTA DL approaches:}
In Table \ref{tab:concatcnn_comparison}, we compare our ViT-based reconstruction approach with the results of Concatenate Convolutional Neural Network from the paper \cite{Lee2021UsingDL}, which claims to achieve the best performance among the different DL models. We present results on the MIT FiveK dataset using a 5x5 defect cluster in the center of a 15x15 patch, similar to \cite{Lee2021UsingDL}. Since the defect cluster location is known in advance according to the assumption in \cite{Lee2021UsingDL}, we divide the image into 5x5 patches and mask the defective patch embedding in the center before sending the embeddings into the encoder. Thus, the value of the defective patch in the center is predicted based on the neighboring 8 patches. Since the input size is only 15x15, we use only 2 encoder and 2 decoder layers for the ViT model with an embedding length of 16. Remarkably, our method attains similar performance with less than half the parameters and computational cost and does not need the location of the bad pixels. In particular, if we do not know the bad pixel locations, we can simply run our model on all image patches independently, reconstructing the entire image. 

\begin{table}[htbp]
    \centering
    \begin{tabular}{c|c|c|c}
    \toprule
        \textbf{Method} & \textbf{NMSE} & \textbf{Params} & \textbf{FLOPs} \\
        \midrule
        Concat CNN \cite{Lee2021UsingDL} & 0.003 & 26.84 K & 203.89 KMac \\
        \midrule
        ViT AE (Ours) & 0.004 & 11.36 K & 102.3 KMac \\
        \bottomrule
    \end{tabular}
    \vspace{2mm}
    \caption{Comparison of ViT-based reconstruction vs pixel correction using Concatenate CNNs \cite{Lee2021UsingDL} for a $5\times5$ defect cluster}
    \label{tab:concatcnn_comparison}
\end{table}

\subsection{Ablation Studies}
\begin{figure}[htbp]
\begin{minipage}[b]{.48\linewidth}
  \centering
  \centerline{\includegraphics[width=0.9\linewidth]{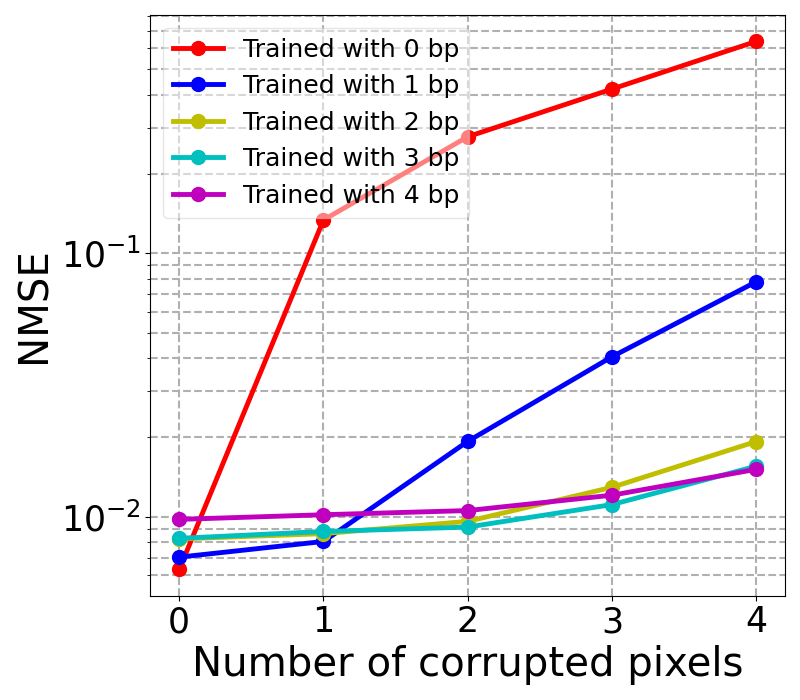}}
   \label{fig:separate_models}
  \centerline{(a) Training with corrupted pixels}\medskip
\end{minipage}
\hfill
\begin{minipage}[b]{0.48\linewidth}
  \centering
  \centerline{\includegraphics[width=0.9\linewidth]{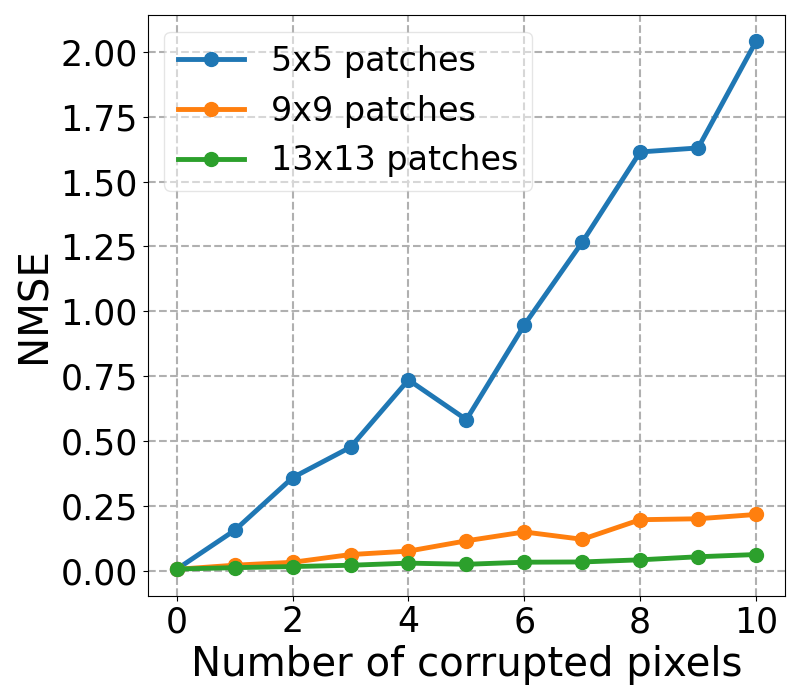}}
       \label{fig:patch_size}
  \centerline{(b) Increasing patch size}\medskip
  
\end{minipage}
\vspace{-0.5cm}
\caption{Patch-based pixel correction on the S7 ISP dataset}
\label{fig:corr_patch}
\end{figure}

\noindent\textbf{Necessity for Training with Corrupted Pixels:}
For patch-based pixel correction using MLP (Figure \ref{fig:mae_vs_mlp}), we assess if models need to be trained with bad pixels in the neighborhood. In Figure \ref{fig:corr_patch}(a), we present results for models trained with patches at different error rates. We observe that no single model achieves the best performance for all error rates. A model trained using no bad pixels in the neighborhood performs poorly when it is fed with a patch containing up to four bad pixels. On the other hand, a model trained with four bad pixels incurs a small increase in error for cases with no neighborhood defects, although it yields relatively low error for all defect rates. Hence, we train separate models for a given error rate.

\vspace{2mm}
\noindent\textbf{Impact of Increasing Patch Size:} In Figure \ref{fig:corr_patch}(b), we demonstrate results with an increased patch size of $9\times 9$ and $13\times 13$ when there are up to 10 bad pixels in the neighborhood. The models are trained on patches with no bad pixels in the neighborhood and tested on patches with multiple bad pixels. Increasing patch size provides a clear advantage. 

\section{Summary and Conclusions}
\label{sec:conclusion}
This paper presents novel and comprehensive DL based solutions for both the detection and correction of bad pixels for image sensors, for a wide range of error rates, and pixel variations. We achieve detection rate up to 99.6\% with less than 0.6\% false positives.  
The correction algorithm yields significantly better results than classical interpolation based approaches. We also offer a fail-safe reconstruction approach for extremely high error rates, which achieves 1.55\% average pixel error for 70\% corrupted pixels.
Our future work includes exploring how the correction algorithm can be combined with in-sensor computing solutions.

Since our pixel detection and correction pipeline operates on the pre-ISP raw images, 
it can also completely bypass the ISP operations \cite{Datta2022EnablingIL}, which are typically expensive and performed off-chip. This also enables the pathway for our pipeline to be integrated with existing in-pixel computing paradigms \cite{Datta2022APP,datta2023icassp},
that can significantly improve the sensor energy efficiency for CV tasks.  

\subsubsection{Acknowledgements} 
We thank Dr. Souvik Kundu for his guidance in this research.

%
%
%
\bibliographystyle{splncs04}
\bibliography{refs}
%




\end{document}